\documentclass[floatfix,11pt,onecolumn,showpacs,amsmath,amssymb]{revtex4}

\usepackage{epsf}
\usepackage{graphicx}  
\usepackage{dcolumn}   
\usepackage{bm}        

\newcommand{\be}{\begin{equation}}
\newcommand{\en}{\end{equation}}
 \newcommand{\bea}{\begin{eqnarray}}
 \newcommand{\ena}{\end{eqnarray}}
  \newcommand{\sch}{Schwarzschild}

\begin{document}

\title{Brown-York energy in stationary spacetimes}
\author{Hongsheng Zhang$^{1,2~}$\footnote{Electronic address: sps\_zhanghs@ujn.edu.cn}}
\affiliation{ $^1$ School of Physics and Technology, University of Jinan, \\336 West Road of Nan Xinzhuang, Jinan, Shandong 250022, China\\
$^2$ State Key Laboratory of Theoretical Physics, Institute of Theoretical Physics, Chinese Academy of Sciences, Beijing, 100190, China}

\date{ \today}

\begin{abstract}
 One usually defines the Brown-York energy for a 2-surface
embedded in a spacelike 3-slice as an integration of  the mean curvature of the 2-surface isometrically embedded
into the 3-slice, with a proper reference 3-space. We demonstrate that this naive definition is ill for stationary
 spacetimes. As an example, we investigate the Kerr-Newman spacetime in detail. We show that the naive definition of the Brown-York energy
 is not a component of the Brown-York boundary stress tensor,  and thus deviates from the original idea of Brown and York. Furthermore, we
 present the exact form of the Brown-York energy for the Kerr-Newman spacetime with the proper reference.

\end{abstract}

\keywords{Brown-York energy; stationary spacetimes}
\preprint{arXiv: }
 \maketitle

\section{Introduction}

    The Einstein field equation describes the gravitational fields yielded by some stress-energy distributions of matters. Surprisingly, the stress-energy, especially the energy (mass) of gravitational fields is an intricate problem. Gravitational fields must be associated energy. In fact we have a well-defined  total energy on an asymptotically flat manifold, that is, the Arnowitt-Deser-Misner (ADM) energy \cite{adm}. But a total energy is not enough. In the problems of gravitational collapse, we need study how much energy is transferred from matter fields to gravitational fields. Note that the energy of matter fields can be localized, so one has energy density of a matter field. In the observations of gravitational waves, a wave may propagate a billion years to arrive at the earth. We cannot say that a gravitational wave carries energy from the binary star to us if the energy of the gravitational field can only be defined globally. A wave is non-physical if it does not carry energy. However, the energy of gravitational fields is non-localizable because of the equivalence principle. As an inevitable compromise, people consider the quasi-local energy of gravitational fields, i.e., the energy enclosed in a finite 2-surface.

    The positive energy theorem says that the ADM energy is larger than or equal to zero for a manifold satisfying some energy conditions (without black holes on it) \cite{yao1}. Later, some other types of total energy, such as Bondi-Sachs energy is also shown to be positive definite or zero \cite{bondi}. The positive energy theorem is profound since it ensures that the Minkowski space is stable. Otherwise the Minkowski space may decay to some other curved configurations. The Penrose conjecture is a natural generalization of the positive energy theorem, in which a quasi-local energy is involved \cite{pen}. Numerical relativity plays a central role to generate the templates of gravitational waves of the compact binaries \cite{ligo}. In numerical calculations, all the domains involved are finite, and thus the quantities in such domains must be quasi-local.

    We still do not reach a consensus about the quasi-local energy of the gravitational field. Several different quasi-local energies have been proposed, for a review stressing on the physical ideas of different quasi-local energies, see \cite{review}. Among all these forms, the Brown-York expression is a leading one \cite{BR}. The physical idea of Brown-York mass originates from an observation of the classical Hamilton-Jacobi equation, that is, the energy of a system can be expressed as a derivative of the action with respect to time on the boundary. As an extension of this idea to the gravity theory, Brown and York define the gravitational boundary stress and a variation of the action with resect to the boundary metric. Thus the zero-zero component of boundary stress should be the gravitational surface energy density. An integration leads to the gravitational energy enclosed in the 2-surface. Such an energy usually diverges for an infinite surface. One needs to introduce a reference metric to counter this divergence, such that one can obtain a finite result. In some cases, it is difficult to find proper reference space and reference metric even if they satisfy the embedding theorems. For example, only asymptotic form of the reference metric at large $r$ are obtained for the Kerr-Newman metric \cite{kerr1}. The exact form is still in absence. The crucial flaw in the previous calculations of Brown-York mass of the Kerr-Newman metric is that the mean curvature of a 2-surface isometrically embedded into a 3-slice with $t$=constant is treated as the surface energy density. One will see that this definition has little relation to the physical problems in rotating spaces. This problematic definition of Brown-York mass appeared in many previous references \cite{many}.

    The boundary Brown-York stress is extensively applied in AdS/CFT \cite{adscft}, which is one of the most significant progress in theoretical physics in recent years. To obtain a concrete proper form of the Brown-York stress is urgent and essential to these problems. In the next section, we present the concrete form of Brown-York energy in stationary spacetimes with proper reference metric. In section III, we study the properties of this energy form. We demonstrate it has the desired properties. For example it reduces to the ADM energy when $r\to \infty$, larger than $m$ but less than $2m$ at the horizon. In section IV, we present our conclusion.

   \section{Brown-York energy in stationary spacetime}
   We start from a general axial symmetric spacetime,
   \be
   ds^2=g_{00}(t,r,\theta)dt^2+2g_{03}(t,r,\theta)dtd\phi+g_{11}(t,r,\theta)dr^2+g_{22}(t,r,\theta)d\theta^2+g_{33}(t,r,\theta)d\phi^2,
   \label{fmetric}
   \en
   where $t,~r,~\theta,~\phi$ are four coordinates on the 4-manifold, and $g_{\mu\nu}$ are components of the metric $g$ of this manifold. According to the definition, $dr$ is the normal covector of the hypersurface $r=$constant. Thus we obtain the normalized normal vector $n$ of this hypersurface,
   \be
    n^a=\alpha g^{ab}(dr)_{b},
    \en
   where $\alpha$ is a normalization factor. Requiring $n_an^a=1$, we reach
    \be
    n^a=g_{11}^{-1/2}\left(\frac{\partial}{\partial r}\right)^a.
    \en
   According to definition, the Brown-York stress is the variation of the boundary term with respect to the boundary metric,
   \be
   8\pi\tau_{\mu\nu}=-\left(K_{\mu\nu}-K\gamma_{\mu\nu}\right),
   \label{tau00}
   \en
   where $K_{\mu\nu}$ is the second fundamental form of the hypersurface $r=$constant, $K=K_{\mu\nu}g^{\mu\nu}$, and $\gamma$ is the induced boundary metric,
   \be
   \gamma^{ab}=g^{ab}-n^an^b(g_{cd}n^cn^d).
   \en
   Eq.(\ref{tau00}) implies,
   \be
   8\pi\tau_{0}^0=-(K_{0}^0-(K_0^0+K_2^2+K_3^3))=K_2^2+K_3^3,
   \label{K23}
   \en
   where we have used $K_1^1=0$. Eq.(\ref{K23}) shows that the boundary energy density equals a sum of the last two components of the second fundamental form of the hypersurface. However, this can $not$ be further simplified to the mean curvature of the 2-surface isometrically embedded
into the 3-slice with $t=$constant, which is widely accepted as the definition of the Brown-York energy density.  We demonstrate this point in the spacetime (\ref{fmetric}). Directly calculation presents,
  \be
  K_2^2=\frac{g_{22}'}{2g_{22}\sqrt{g_{11}}},
  \label{K22}
  \en
  and
  \be
  K_3^3=\frac{g_{30}g'_{30}-g_{00}g_{33}'}{2(g_{30}^2-g_{00}g_{33})\sqrt{g_{11}}},
  \label{K33}
  \en
  where a prime denotes a derivative with resect to $r$.
  As a contrast, we calculate the mean curvature of the 2-surface $r=$constant imbedded in a 3-slice with $t=$constant. First the induced metric of this 3-slice is,
  \be
   ds^2=g_{11}(t,r,\theta)dr^2+g_{22}(t,r,\theta)d\theta^2+g_{33}(t,r,\theta)d\phi^2.
   \label{3metric}
   \en
  The following calculation is simple and straightforward. The components of the second fundamental form of the 2-surface $r=$constant in this 3-slice $k_{\mu\nu}$ read,
   \be
   k_2^2=\frac{g_{22}'}{2g_{22}\sqrt{g_{11}}},
  \label{k22}
  \en
   \be
   k_3^3=\frac{g_{33}'}{2g_{33}\sqrt{g_{11}}},
   \label{k33}
   \en
  and $k_1^1=0$. It is clear $K_2^2=k_2^2$, but $K_3^3\neq k_3^3$. So we have,
  \be
  8\pi\tau_{0}^0=K_2^2+K_3^3\neq k_2^2+k_3^3.
  \en
   This implies that the mean curvature of the 2-surface imbedded in the 3-slice does not equal to the surface energy density, and is thus not a component of the Brown-York stress. Our demonstrations disqualified the mean curvature of such a 2-surface as the surface energy density. To obtain the proper form of the surface energy density, we need to go back to the whole spacetime rather than just starting from a 3-slice.  Only for time-orthogonal spacetimes, $K_2^2+K_3^3$ coincides with $k_2^2+k_3^3$. In such cases, for example, the \sch~spacetime, $k_2^2+k_3^3$ represents the surface energy density. The above observation may lead to more extensive results than what we pointed out in frame of Brown-York energy. In the expressions of Liu-Yao \cite{yao2}, Wang-Yao\cite{yao3}, and Kijowski energies \cite{kijo}, the mean curvature of the 2-surface imbedded in the 3-slice are involved. In analogy to the above discussions, whether they correspond physical energies in rotating spacetimes needs to be studied seriously.

   If $g_{\mu \nu}$ in (\ref{fmetric}) are not functions of $t$, it is a stationary metric. In this case, we can reconstruct the above results in the conservation charge picture. The gravitational energy is not an isolated problem. An ideal way is to define it as a conserved charge of a conserved current, and thus we can define the momentum, and further the angular momentum of the gravitational field in a unified way. For example, the original form of Misner-Sharp energy was presented only as a quasi-local energy of the gravitational field, while the corresponding gravitational momentums are not considered. The subsequent researches show that the Misner-Sharp mass can be treated as the conserved charge of the conserved current in a static spacetime,
   \be
   J_{MS}^a=T^{ab}\xi_b,
   \en
   where $T$ denotes the stress-energy of the spacetime, and $\xi$ is a time-like Killing vector of the spacetime. The Misner-Sharp mass is defined in an elegant way,
   \be
   M_{MS}=\int *J_{MS},
   \en
   where $*$ denotes the Hodge dual of a tensor, which performs in the total spacetime. In this way, the momentum of a gravitational field also gets natural definition. Similarly, the Brown-York energy can be defined as the boundary conserved charge of a boundary conserved current,
   \be
   J_{BR}^a=\tau^{ab}\xi_b,
   \en
      if a Killing vector $\xi=\frac{\partial}{\partial t}$ is permitted at the boundary spacetime with $r=$constant,
      \be
      B_{ab}=g_{00}(r,\theta)dt^2+2g_{03}(r,\theta)dtd\phi+g_{22}(r,\theta)d\theta^2+g_{33}(r,\theta)d\phi^2,
      \label{Bab}
      \en
   the Brown-York energy reads,
   \be
    M_{BR}=\int *J_{BR}.
   \en
   One can show $J_{BR}$ is really a conserved current.
   \be
   \nabla_a J_{BR}^a=\nabla_a \tau^{ab}\xi_b+\tau^{ab}\nabla_a \xi_b.
   \en
   Both of the two terms in the right hand side of the above equation equal zero, since $\tau_{ab}$ is conserved, and $\xi_b$ is Killing at the same time $\tau_{ab}$ is symmetric. We thus derive $\nabla_a J_{BR}^a=0$.
   Note that here the Hodge dual operator performs on the 3-boundary $B_{ab}$ shown in (\ref{Bab}). One calculates $*J_{BR}$ directly,
   \be
   *J_{BR}=-\sqrt{g_{22}g_{33}}\tau^0_0d\theta\wedge d\phi=-\frac{1}{8\pi}\sqrt{g_{22}g_{33}}(K^2_2+K^3_3)d\theta\wedge d\phi.
   \en
   Thus, we exactly obtain the same result using the conserved charge method.

   We apply the above result to the case of the Kerr-Newman metric. The Kerr-Newman metric may be the unique asymptotically flat stationary metric with electromagnetic charge in the Einstein gravity, which is  a special case of the metric (\ref{fmetric}). In the Boyer-Lindquist coordinates, the components of the Kerr-Newman metric read,
    \be
    g_{00}=-(1-\frac{2mr-q^2}{\rho^2}),
    \label{g00}
    \en
    \be
    g_{03}=-\frac{a(2mr-q^2)}{\rho^2}\sin^2\theta,
    \en
    \be
    g_{11}=\frac{\rho^2}{\Delta},
    \label{g11}
    \en
    \be
    g_{22}=\rho^2,
    \label{g22}
    \en
    \be
    g_{33}=\frac{(2mr-q^2)a^2\sin^4\theta}{\rho^2}+(r^2+a^2)\sin^2\theta.
    \label{g33}
    \en
   Here, as usual,
   \be
   \rho^2=r^2+a^2\cos^2\theta,
   \en
   and
   \be
   \Delta=r^2-2mr+a^2+q^2.
      \en
 $m,~q,~a$ are three parameters in the metric, denoting the total mass, total charge, and angular momentum per mass respectively. The corresponding
 electromagnetic potential reads,
 \be
 A=-\frac{qr}{\rho^2}(dt-a\sin^2\theta d\phi).
 \en
    Based on the above discussions one immediately sees that  $K_2^2+K_3^3\neq k_2^2+k_3^3$ for the Kerr-Newman case. The proper surface energy density $\sigma$ reads,
    \be
    \sigma=-{\tau_0^0}=-\frac{K_2^2+K_3^3}{8\pi}.
    \en
    Substituting the metric components of Kerr-Newman into (\ref{K22}) and (\ref{K33}), we obtain the surface density $\sigma$,
    \be
    \sigma=-\frac{1}{8\pi}\frac{ L-a^4 (m-r)\cos 4 \theta +4 a^2 r \cos 2 \theta \left(2 a^2+3 r (r-m)+q^2\right)}{\sqrt{2} \left(a^2 \cos 2 \theta +a^2+2 r^2\right)^{5/2}}\sin \theta \sqrt{\frac{\left(a^2+r^2\right)^2-a^2\Delta \sin ^2\theta }{\Delta}},
    \label{densityKN}
    \en
    where,
    \be
    L=a^4 m+7 a^4 r-20 a^2 m r^2+12 a^2 q^2 r+20 a^2 r^3-32 m r^4+16 q^2 r^3+16 r^5.
    \en
    It is easy to check that the surface density degenerates to
    \be
    \sigma=-\frac{1}{4\pi} \sin \theta  \sqrt{r (r-2 m)},
    \en
     when $m=0,~q=0$, which is exactly the unreferenced Brown-York energy density of the \sch~spacetime.
   All the previous studies on Kerr-Newman sacetime treat $\sigma_w=-(k_2^2+k_3^3)/(8\pi)$ as the surface density. From  (\ref{k22}) and (\ref{k33}), we obtain,
   \be
   \sigma_w=-\frac{1}{8\pi}\frac{r \sin \theta   \left(2 r \left(a^2+r^2\right)-a^2 \sin ^2\theta  (r-m)\right)}{ \sqrt{\left(a^2 \cos ^2\theta +r^2\right) \left(\left(a^2+r^2\right)^2-a^2 \Delta \sin ^2\theta  \right)}}\sqrt{\frac{a^2+q^2}{r^2}-\frac{2 m}{r}+1}.
   \en
   It is easy to check the above density is exactly the results in \cite{kerr1}. Essentially, this surface density corresponds a hypothetical metric whose components equal  to the Kerr-Newman's components besides $g_{03}$,
   \be
   ds^2_i=g_{00}dt^2+g_{11}dr^2+g_{22}d\theta^2+g_{33}d\phi^2,
   \en
   where $g_{00}$, $g_{11}$, $g_{22}$, and $g_{33}$ are given by (\ref{g00}), (\ref{g11}), (\ref{g22}), and (\ref{g33}), respectively. This hypothetical metric is not a solution of the Einstein-Maxwell equation (we name it HMNEM later). One sees that the Kerr-Newman spacetime has a different Brown-York surface energy density compared with this  HMNEW, although they share the same 3-slice. Thus, it is far from enough if one uses a 3-manifold as the starting point when  investigating the Brown-York energy. The Brown-York energy depends on the whole spacetime. This is our principle observation in this letter.

   Then we continue to complete the Brown-York energy in a compact 2-surface with $r=$constant and $t=$constant for the Kerr-Newman spacetime and present the proper reference metric. The energy enclosed in this  2-surface is a simple integral
  \be
    E_{BY}=\int _S \sigma,
    \en
  where $S$ denotes the 2-surface, and $\sigma$ is given by (\ref{densityKN}). The result is a little lengthy,
  \bea
  E_{BY}=\frac{A_0}{6(1+A_1)^{3/2}A_1^{3/2}(A_1-A_3)(1+A_3)^{1/2}} \left(iB_1E\left(i\ln \frac{1+\sqrt{A_1+1}}{\sqrt{A_1}}|\frac{A_1}{A_3}\right)\right. \nonumber
    \\ \left.+\sqrt{\frac{1+A_1}{A_1}}\left(B_2+iB_3F\left(i\ln \frac{1+\sqrt{A_1+1}}{\sqrt{A_1}}|\frac{A_1}{A_3}\right)\right)\right),
 \ena
  where $E$ denotes the elliptic integral of the second kind, which is defined as,
  \be
  E(x|y)=\int_0^x(1-y\sin^2\theta)^{1/2}d\theta,
  \en
  and  $F$  denotes the elliptic integral of the first kind, which is defined as,
  \be
  F(x|y)=\int_0^x(1-y\sin^2\theta)^{-1/2}d\theta.
  \en
   The variables are defined as follows,
  \be
  B_1=(1+A_1^{-1})^{1/2}(1+A_3^{-1})^{1/2}(1+A_1)A_3\left(2A_1^2+A_1(A_2-A_3)-2A_2A_3\right),
  \en
  \be
  B_2=(1+A_1)^{-1/2}(1+A_3)(A_1^3+2A_1^2(1+A_2)-2A_2A_3+A_1(A_2-A_3-3A_2A_3)),
  \en
  \be
  B_3=A_3(1+A_1)(A_1+2A_2)(1+A_3^{-1})^{1/2}(A_3-A_1).
  \en
  Here $A_0,~A_1,~A_2,~A_3$ are defined by,
  \be
  A_0=\frac{a^2 (m-5 r)-2 r \left(3 r (r-m)+q^2\right)}{2 a^2 },
  \en
  \be
  A_1=\frac{r^2}{a^2},
  \en
  \be
  A_2=\frac{a^4 (r-m)-2 a^2 r \left(r (r-m)+q^2\right)-4 r^3 \left(r (r-2 m)+q^2\right)}{a^2 \left(a^2 (m-5 r)-2 r \left(3 r (r-m)+q^2\right)\right)},
  \en
  \be
  A_3=\frac{(a^2+r^2)^2}{\Delta a^2}-1.
  \en
  When $a\to 0$, the Brown-York energy of Kerr-Newman degenerates to
  \be
  E_{BY}=-\sqrt{r^2-2mr+q^2}.
  \en
  This is exactly the unreferenced Brown-York energy of the Reissner-Nordstrom black hole.

  A proper reference metric is critical to obtain the exact form of the Brown-York energy. For the \sch~and  Reissner-Nordstrom spacetimes, a Minkowski space is the operable reference metric. However, even for the HMNEW, a suitable reference metric is still does not found. Here we find the proper reference metric for the Kerr-Newman spacetime as follows,
  \be
  ds^2=-dt^2+\frac{\rho^2}{r^2+a^2}dr^2+\rho^2d\theta^2+(r^2+a^2)\sin^2\theta d\phi^2.
  \label{BY0}
  \en
  It is nothing but the Minkowski metric in the ellipsoidal coordinates. In principle,
  the reference space should be Minkowskian one in ellipsoidal coordinates, since here we write the Kerr-Newman metric in quasi-ellipsoidal coordinates (Boyer-Lindquist). The Boyer-Lindquist coordinates reduce to the ellipsoidal ones when $M=0$ and $q=0$ for Kerr-Newman. We thus deduce that the proper reference is the  Minkowski space in ellipsoidal coordinates. When one needs the Brown-York energy in other coordinates, for example, in the  Kerr-Schild Cartesian coordinates, we should use a reference as the Minkowski in Cartesian coordinates. In this sense, the reference metric is covariant.

  Now we discuss the Brown-York energy in this reference metric. As we have analysed in detail in the previous discussions, in this case (\ref{BY0}) $K_2^2+K_3^3$ equals $k_2^2+k_3^3$, since it is a time-orthogonal space. Thus the surface energy density reads,
  \be
  8\pi \sigma= -(K_2^2+K_3^3)=- (k_2^2+k_3^3).
  \en
  From (10) and (11), we obtain,
  \be
  \sigma=-\frac{1}{8\pi}\left(\frac{g_{22}'}{2g_{22}\sqrt{g_{11}}}+\frac{g_{33}'}{2g_{33}\sqrt{g_{11}}}\right).
  \en
  Substituting the components of (43), we thus arrive at the effective surface energy density in the coordinates (43),
  \be
  \sqrt{g_{22}g_{33}}\sigma=-\frac{1}{8\pi}\frac{(3a^2+4r^4+a^2\cos2\theta)r\sin\theta}{a^2+2r^2+a^2\cos2\theta}.
  \en
 Integrating the density over the 2-space $(\theta,\phi)$, we obtain,
 \be
  E_{BY0}=\int_0^{2\pi}d\phi\int_0^{\pi}d\theta\left(-\frac{1}{8\pi}\frac{(3a^2+4r^4+a^2\cos2\theta)r\sin\theta}{a^2+2r^2+a^2\cos2\theta}\right)=-\frac{r}{2}-\frac{a^2+r^2}{2a}{\arctan}\frac{a}{r}.
 \en
  Then we obtain the exact form of the Brown-York energy enclosed in a 2-surface with $r=$constant and $t$=constant for the Kerr-Newman spacetime,
  \be
  E_{BYKN}=E_{BY}-E_{BY0}.
  \label{ebykn}
  \en
 This equation enables us to calculate the Brown-York energy in analytical form.

  One can also define the Brown-York mass by using the asymptotic behaviour of $E_{BY0}$. Expanding the unreferenced energy around $r\to \infty$,
  \be
   E_{BY0A}=-r-\frac{a^2}{3r}+\frac{a^4}{15r^3}+\mathcal {O}(1/r^5).
   \en
  At the leading order, it is exactly the same as of a \sch~ one. By using this form of the unreferenced Brown-York energy, we obtain,
    \be
  E'_{BYKN}=E_{BY}-E_{BY0}.
  \en
  $E'_{BYKN}$ degenerates to $E_{BYKN}$ in (\ref{ebykn}) when $r\to \infty$.
  
 Let's just recapitulate the essential difference of the definitions of Brown-York energy in previous literatures and in present letter. As we have mentioned, all the previous discussions of the Brown-York energy treated the mean curvature of the 2-space as the surface energy density \cite{kerr1}. It can be written as,
  \be
  \sigma_w=-\frac{k_2^2+k_3^3}{8\pi}.
   \en
   By using (10) and (11), we obtain,
   \be
   \sigma_w=-\frac{1}{8\pi}\left(\frac{g_{22}'}{2g_{22}\sqrt{g_{11}}}+\frac{g_{33}'}{2g_{33}\sqrt{g_{11}}}\right).
    \en
    Thus, the BY energy enclosed in such a 2-surface reads,
    \be
    E_{BYw}=\int_S \sigma_w=-\int_0^{2\pi}d\phi\int_0^{\pi}d\theta  \sqrt{g_{22}g_{33}}\frac{1}{8\pi}\left(\frac{g_{22}'}{2g_{22}\sqrt{g_{11}}}+\frac{g_{33}'}{2g_{33}\sqrt{g_{11}}}\right).
    \en
    In the above expression, neither $g_{00}$ nor $g_{03}$ appear. To obtain the energy, only information of the metric of the 3-slice is necessary.
    As comparison, in our expression, the BY energy reads,
    \be
    E_{BY}=-\int_0^{2\pi}d\phi\int_0^{\pi}d\theta  \sqrt{g_{22}g_{33}}\frac{1}{8\pi}\left(\frac{g_{22}'}{2g_{22}\sqrt{g_{11}}}+\frac{g_{30}g'_{30}-g_{00}g_{33}'}{2(g_{30}^2-g_{00}g_{33})\sqrt{g_{11}}}\right).
    \en
    Clearly, one sees that all the components of the full metric are involved, and especially the essential character of the stationary space $g_{03}$ is involved. But $E_{BYw}$ does not consider the rotation effects of spacetime. This is to say, according to the previous definition in literatures of BY energy, spacetime with same spacelike 3-slice,
    \be
    dl^2=g_{11}dr^2+g_{22}d\theta^2+g_{33}d\phi^2,
    \en
    but different rotations even without rotation will share the same gravity energy. We think this is not reasonable.
    In our refined definition, the BY energy needs the information of the full metric. We think this is more reasonable.

    This is an abstract discussion of the two definitions, i.e., we don't require the metric is an on-shell configuration of some gravity theory. For a concrete gravity theory, for example the Einstein gravity, the field equation imposes relations among $g_{\mu\nu}$. Thus they are not independent. In this sense, $g_{11}, g_{22}, g_{33}$ can carry the information of the full metric.

    \section{Properties of the Brown-York energy in stationary spacetimes}
    Because the expression of Brown-York energy in (34) is rather lengthy. We try to show some limits to investigate its asymptotic behaviour. And one will see that it has all the desired properties. One can obtain the familiar results in time orthogonal spacetimes. First, when $q=0$ and $m=0$ (34) degenerates to
   \be
   E_{BY}= -\frac{r}{2}-\frac{a^2+r^2}{2a}{\arctan}\frac{a}{r}.
   \en
   That is exactly $E_{BY0}$. The outer horizon of Kerr-Newman reads,
   \be
   r_h=m+\sqrt{m^2-a^2-q^2}.
   \en
   At the horizon we have
   \be
   \lim_{r\to r_h} E_{BY}=0.
   \en
   Still this property is the same as the \sch~black hole. Thus the horizon mass of Kerr-Newman black hole with reference spacetime,
   \be
   E_{HKN}= \frac{r_h}{2}+\frac{a^2+r_h^2}{2a}{\arctan}\frac{a}{r_h}.
   \label{EEH}
   \en
   We find that,
   \be
   M<E_{HKN}\leq 2M.
   \en
   When $q=a=0$, it reduces to the case of the \sch~, and thus the equality holds. The extreme black hole is an interesting topic. A \sch~black hole cannot be extreme. Here we explore Brown-York energy
   for extreme Kerr-Newman black holes. For an extreme black hole, the horizon mass reads
   \be
   E_{HKNE}= \lim_{r_h \to M}\left(\frac{r_h}{2}+\frac{a^2+r_h^2}{2a}{\arctan}\frac{a}{r_h}\right).
   \en
   $E_{HKNE}\leq M$, and the equality holds iff $a=0$.

   Then we explore its behaviour at large distance by two plots.
   \begin{figure}
$\begin{array}{cc}
\includegraphics[width=0.45\textwidth]{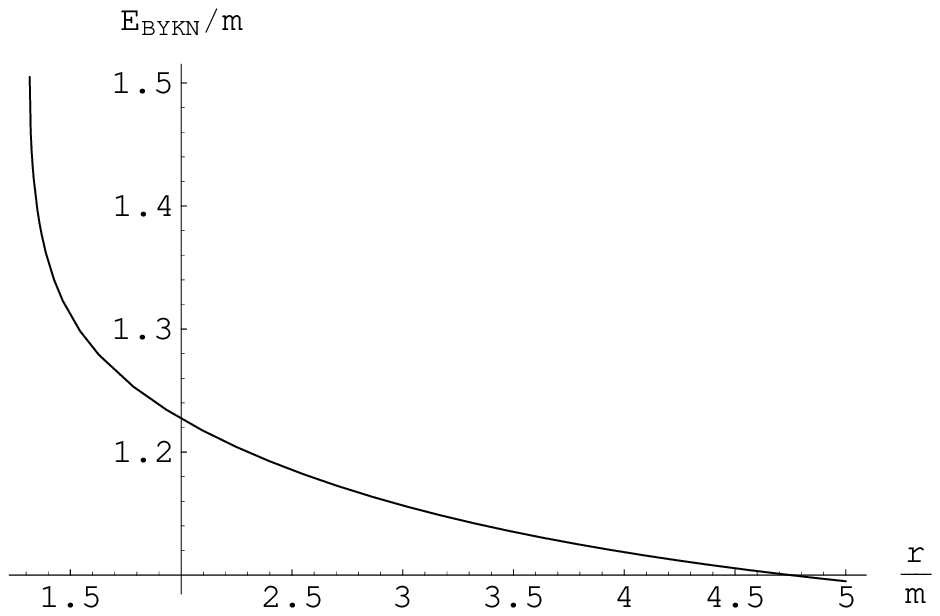}&
\includegraphics[width=0.45\textwidth]{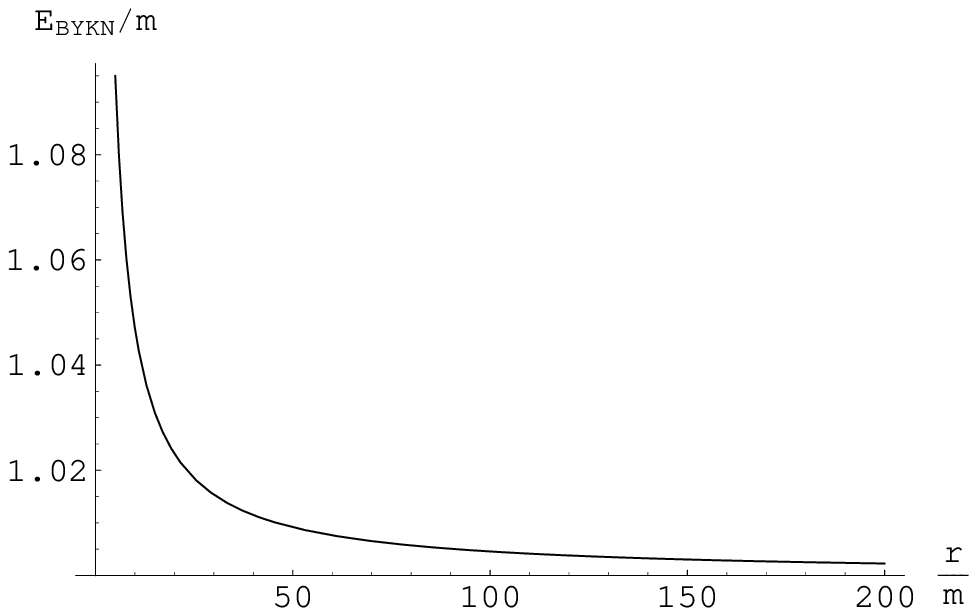}
\end{array}$
 \caption{In this figure, we show the Brown-York energy of a Kerr-Newman black hole with $a/m=0.9,~q/m=0.3$. This figure displays the behaviour of referenced Brown-York energy $E_{BYKN}$ from the horizon to some large distance. Note that $E_{BYKN}$ is not divergent at the horizon, as shown in (\ref{EEH}). Here it is 1.506m.}
 \label{X1}
 \end{figure}
   From the figure (\ref{X1}), one finds that the referenced Brown-York energy has a nice property. It converges to the ADM energy at large distance, which is never obtained in the previous studies.

   \section{conclusion}
 Now we present a short summary of this letter. The problem of gravitational energy is an important problem in both theoretical and
 practical aspects. Unexpectedly, the gravitational is essentially nonlocal even at the classical level. However, it is impossible to make out the energy loss from a binary bodies in the gravitational radiation problems if one only has a global definition of gravitational energy. Thus several quasilocal forms of gravitational energy are developed. Based on the Hamilton-Jaccobi approach, the Brown-York energy is presented. Traditionally, the Brown-York energy enclosed in  a 2-surface embedded in a spacelike 3-slice is calculated by an integration of  the mean curvature of the 2-surface isometrically embedded into the 3-slice, with substraction by a proper reference 3-space (usually a flat space). We find that the mean curvature of the 2-surface does not correspond to the surface energy density for the stationary spacetimes. And thus it is improper to calculate the Brown-York energy through an integration of the mean curvature of a 2-surface isometrically embedded into a 3-slice. In the problems of the Brown-York energy to start from a 3-slice is misleading. The proper Brown-York energy depends on the total spacetime. We demonstrate this point  in a general axial symmetric spacetime, and confirm it by the conserved charge method if there exists a time-like Killing vector. Applying our result to the Kerr-Newman spacetime, we obtain the analytical form of the Brown-York energy for the Kerr-Newman spacetime, which is different from the traditional result \cite{kerr1}. Similar problems in Liu-Yao, Wang-Yao, and Kijowski energies deserve to study further. We conclude that this refined form is more appropriate to realize the original physical idea of Brown and York, i.e., the "00"-component of the boundary stress-energy. And thus it should be as the proper Brown-York energy in  stationary spacetimes. We suggest to use this refined form in the related progresses, such as the proof of positive energy theorem, and AdS/CFT. When a cosmological constant is introduced, we should still treat the
 "00"-component of the boundary stress-energy as the proper Brown-York energy, rather than the mean curvature of a two-sphere.

 {\bf Acknowledgments.}
   This work is supported in part by the National Natural Science Foundation of China (NSFC) under grant Nos. 11575083, and  Shandong Province Natural Science Foundation under grant No.  ZR201709220395.

\end{document}